\documentclass[twocolumn]{aastex63}
\usepackage{graphicx}
\usepackage{amsmath}   
\usepackage{amssymb}

\usepackage{graphicx}  
\usepackage{amsmath}   
\usepackage{amssymb}  
\usepackage{color}

\usepackage{natbib}

\usepackage[active]{srcltx}

\def\eq#1{{Eq.~(\ref{#1})}}

\begin{document}

\title{Constraining the AGN luminosity function from JWST with the X-ray Background}

\author{Hamsa Padmanabhan}
\affiliation{ D\'epartement de Physique Th\'eorique, Universit\'e de Gen\`eve \\
24 quai Ernest-Ansermet, CH 1211 Gen\`eve 4, Switzerland\\
}

\author{Abraham Loeb}
\affiliation{Astronomy department, Harvard University \\
60 Garden Street, Cambridge, MA 02138, USA}

\email{hamsa.padmanabhan@unige.ch, aloeb@cfa.harvard.edu}

 %% Give the title, author etc. here; the format can change depending on the journal
 %% styles. 

  \begin{abstract}
  We predict the X-ray background (XRB) expected  from the population of quasars detected by the JWST spectroscopic surveys over the redshift range $z \sim 4-7$. We find that the measured { UV emissivities} imply a $\sim 10$ times higher unresolved X-ray background than constrained by current experiments. We illustrate the difficulty of simultaneously matching the faint-end of the quasar luminosity function and the X-ray background constraints. We discuss possible origins and consequences of this discrepancy.
 \end{abstract}
 

\section{Introduction}
The recent \textit{James Webb Space Telescope} (JWST) photometric and spectroscopic surveys have discovered a large number of active galactic nuclei (AGN) candidates over the redshifts $z \sim 4.5 - 7$, especially helping to bridge the gap of missing faint AGN with $ M_{\rm UV} < -25$ \citep{matsuoka2023}. The AGN candidates are identified based on X-ray, broad $H\beta$, and high ionization lines. A large number of faint AGN have also been identified at $z \sim 4-7$  \citep[e.g.,][]{onoue2023, kocevski2023, ubler2023, harikane2023, maiolino2023, matthee2023}, with steeper faint ends in their luminosity functions than expected from extrapolating previous results. Such candidates are also detected out to higher redshifts, $z \sim 9-12$ \citep{fudamoto2022,larson2023, goulding2023, maiolino2023a}. Recently, the EPOCHS survey covering the JWST PEARLS and Early Release Observations (ERO) fields identified nine new AGN in the $6.5 < z < 12$ redshift range \citep{ignas2023}, and the UNCOVER survey used the JWST NIRSpec to provide the UV luminosity functions from 15 AGN candidates over $z \sim 4-8$ \citep{greene2023}. The UV luminosity functions (UV LFs) measured at wavelengths 912 \AA\ and 1450 \AA\ are about factors of $\sim 10-100$ greater than the faintest UV-selected quasars observed previously.

The measured bolometric luminosities of the AGN are used to derive the properties of their central supermassive black holes \citep[e.g.,][]{pacucci2023, maiolino2023, larson2023, kocevski2023, natarajan2023}. AGN at $z \sim 4-7$ identified in JWST and previous literature have bolometric luminosities covering a wide range and inferred black hole masses of $10^6 - 10^9 M_{\odot}$.
Many of the derived black-hole to stellar mass ratios are about 10 to a 100 times greater than expected from the local population \citep[e.g.,][]{pacucci2023, yue2023, stone2023}, with one candidate black hole \citep{bogdan2023} also showing evidence for being as massive as the stellar host. 

A population of black holes at high redshifts would contribute to a copious production of X-ray photons with energies $E > E_{\rm max} = 1.8[(1+z)/15]^{0.5} x_{\rm HI}^{1/3}$, where $x_{\rm HI}$ is the neutral hydrogen fraction in the intergalactic medium \citep[IGM;][]{dijkstra2004}.  Here, $E_{\max}$ represents the energy above which the IGM becomes optically thin to photons over a Hubble length. Following the redshifting of these photons without absorption, they would be observed today as the present-day X-ray background (XRB) in the relevant energy regime  \citep[measured with, e.g., the {\it Swift}, INTEGRAL and {\it Chandra} surveys; e.g., ][]{ajello2008, moretti2012, churazov2007, cappelluti2017}.
Previous theoretical and observational studies \citep{shen2020, haiman1999, dijkstra2004, fauchergiguere2020} have found that the current constraints on the quasar population saturate the unresolved component of the measured  XRB \citep{moretti2012, cappelluti2017}. 

In this paper, we predict the level of the XRB implied by the measured AGN UV LFs from the recent JWST surveys over $z \sim 4-7$.  We find that the JWST measurements -- { used in combination with recent findings connecting the UV emissivity to the XRB} --  would oversaturate the {\it Swift} XRT/{\it Chandra} constraints on the unresolved XRB by about an order of magnitude. In so doing, we  illustrate the difficulty of simultaneously matching both the UV LF and the XRB constraints. We discuss the possible reasons and consequences for the discrepancy.

\section{Relevant equations}
The data compilation of \citet{shen2020} allows for a characterization of the bolometric quasar luminosity function from observations in the rest-frame IR, B, UV, soft and hard X-ray wave bands, covering redshifts $z \sim 0-7$. A best-fitting quasar spectral energy density (SED) model is constructed to convert the luminosities across various bands and fitted to the data. The bolometric quasar luminosity function, describing the comoving number density of quasars as a function of luminosity over all wavebands, is found to be well-modeled by a double-power-law \citep{shen2020} form. The same functional form is found to describe the luminosity function in specific wave bands (of rest frequencies $\nu$ with corresponding luminosities $L_{\nu}$ and number densities $n$):
\begin{equation}
\phi (L_{\nu}, z) \equiv \frac{dn}{d L_{\nu}} = \left(\frac{\phi^*_L}{L_*}\right) \left[\left(\frac{L_{\nu}}{L_*}\right)
^{-\gamma_1} + \left(\frac{L_{\nu}}{L_*}\right)^{-\gamma_2}\right]^{-1}
\end{equation}
with the best-fitting parameters $L_*, \gamma_1, \gamma_2$ and $\phi_*$ being matched to observations, both separately at each redshift and globally across the range of redshifts considered.
 The best-fitting form of the luminosity function for the UV (ultraviolet) regime (with rest wavelength 912 \AA) over $z \sim 5-9$ is found to be in good agreement with multiband observational data.
The total emissivity of the AGN in this regime at a given redshift is found by integrating over the luminosity function above a minimum value $L_{\rm min}$:
\begin{equation}
\epsilon_{\nu}(z) = \int_{L_{\nu, \rm min}}^{\infty} L_{\nu} \phi(L_{\nu}, z) \ d L_{\nu}
\end{equation}
and has units of ergs s $^{-1}$ Hz$^{-1}$ Mpc$^{-3}$. 

Using the best-fitting SED template of the AGN emission, the above emissivity can be converted from the UV regime into the X-ray wave band(s). { This can be done in various ways, e.g., in \citet{shen2020}, it is done by normalizing a power-law X-ray SED  template over the soft (0.5-2 keV) and hard (2-10 keV) X-ray regimes using the nonlinear $\alpha_{\rm ox}$ conversion factor}. From  this, the comoving X-ray  emissivity can be computed as a function of redshift $z$ and used to define the cosmic XRB \citep{shen2020, fauchergiguere2020}
\begin{equation}
I_{\rm XRB} = \int_{z_{\rm min}}^{z_{\rm max}} dz \frac{\epsilon_{\nu} (\nu_{\rm em}, z)}{4 \pi d_L^2(z)} \frac{dV}{d \Omega dz}(z)
\label{ixrb}
\end{equation}
with units erg (or keV) cm$^{-2}$ s$^{-1}$ sr$^{-1}$ Hz$^{-1}$, and in which $z_{\rm min}$ and $z_{\rm max}$ are the minimum and maximum redshifts under consideration. In the above, the differential comoving volume element is denoted by
 \begin{equation}
\frac{dV}{d \Omega dz} = \frac{c}{H(z)}(1+z)^2 d_A(z)^2
\label{reddep}
\end{equation}
and $d_A(z)$, $d_L(z)$ are the angular diameter and luminosity distances respectively to redshift $z$. The overall redshift-dependent factor in \eq{ixrb} thus reduces to $c/[H(z)(1+z)^2]$.
The $I_{\rm XRB}$ can equivalently be expressed  as a function of the energy $E$ of emission corresponding to the  rest frequency $\nu_{\rm em}$. The unresolved component of the above background, amounting to about 8-9\% \citep{cappelluti2017} is constrained to originate from sources above $z \sim 4$ \citep{haardt2015}.

\begin{figure}
\begin{center}
\includegraphics[width = \columnwidth]{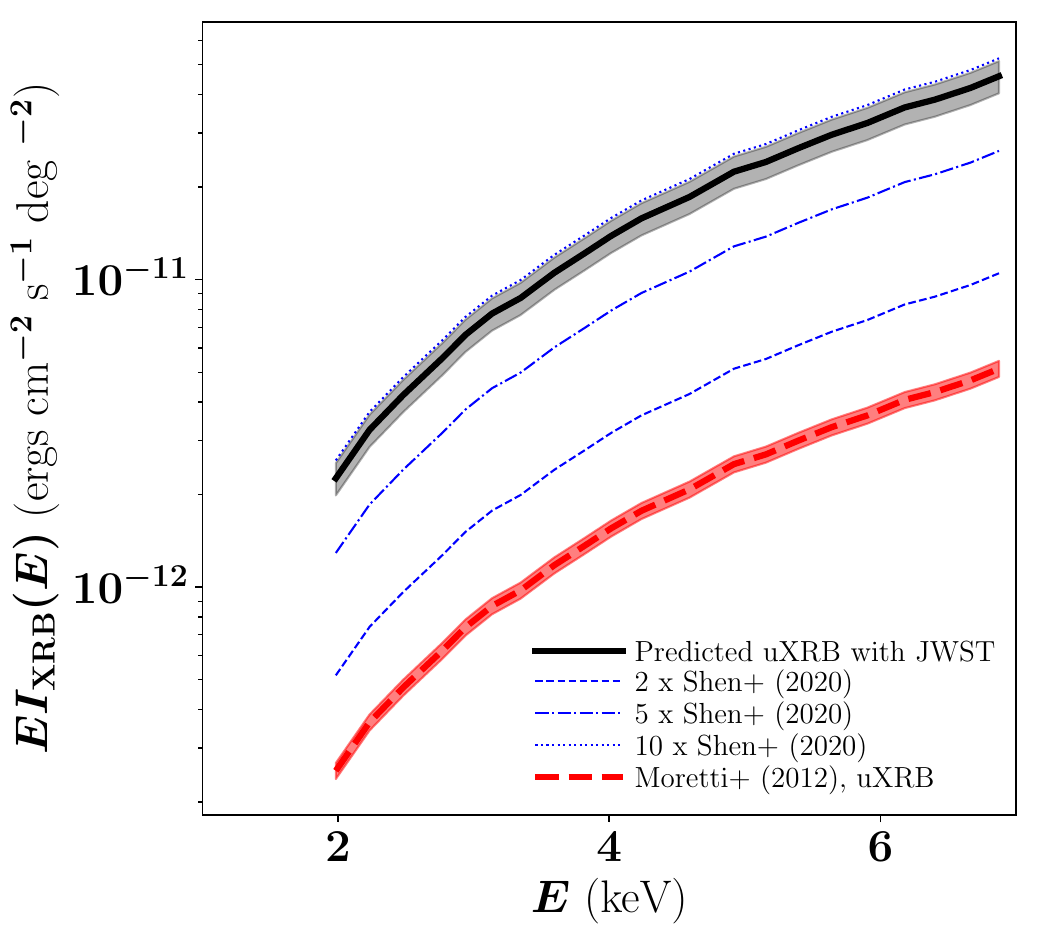}
\end{center}
\caption{Unresolved X-ray background constraints ($E I_{\rm XRB})$ as a function of rest energy $E$ from  fitting to the {Swift} XRT and {Chandra} observations by \citet[][red dashed line]{moretti2012}, { with the red shaded region showing the unresolved XRB measurement uncertainty}. Overplotted are the expected X-ray background from the measured JWST UV LFs (black solid line { and associated gray region}), and those from enhancing the UV emissivities in \citet{shen2020} by factors of $2 \times$, $5 \times$ and $10 \times$ (blue dashed, blue dashed-dotted, and blue dotted lines) over the $z \sim 4-7$ interval.}
\label{fig:xrb}
\end{figure}

The above formalism can be compared to recent measurements of the XRB \citep[e.g.,][]{cappelluti2017}, specifically its unresolved component which is constrained from the { Swift} XRT and {Chandra} data \citep{moretti2012}. This is plotted as the red dashed line in Fig. \ref{fig:xrb}, { with the shaded region representing the uncertainty in the latest unresolved XRB measurements, of the order of 5.5 - 6\%  \citep{cappelluti2017}.}

\section{X-Ray Background from  JWST measurements  of AGN LF}

\begin{figure}
\begin{center}
\includegraphics[width = \columnwidth]{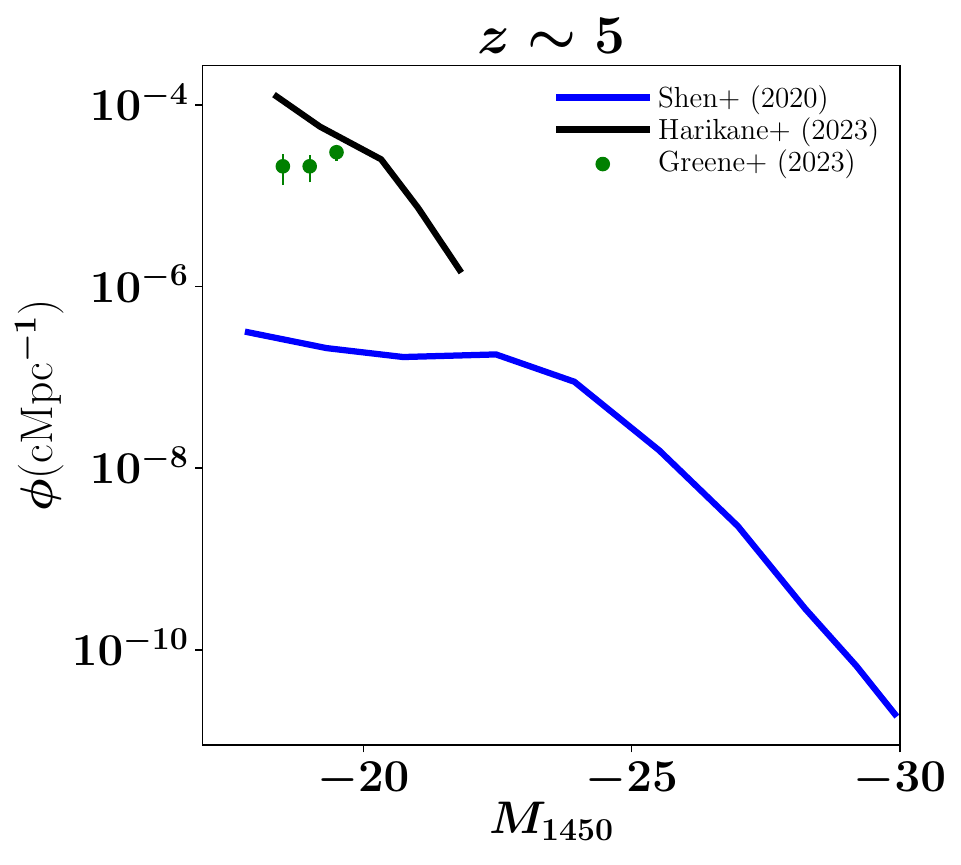}
\end{center}
\caption{The UV LF at 1450 \AA\ at $z \sim 5$ from the results of JWST spectroscopic surveys \citep{harikane2023} and the UNCOVER survey \citep{greene2023}, as compared to the fitting form of \citet{shen2020} matched to previous AGN data. }
\label{fig:uvlf}
\end{figure}

The quasar UV LF at $z \sim 4-11$ compiled from the JWST ERO, PEARLS and CEERS surveys  \citep[][]{matthee2023, maiolino2023, harikane2023, greene2023}  all indicate faint-end values of about a factor of 10 above those expected from extrapolating the fits in the literature (as shown in Fig. \ref{fig:uvlf} for the case of the UV LF measured at rest wavelength 1450 \AA\  at $z \sim 5$). 

\begin{figure}
\begin{center}
\includegraphics[width = \columnwidth]{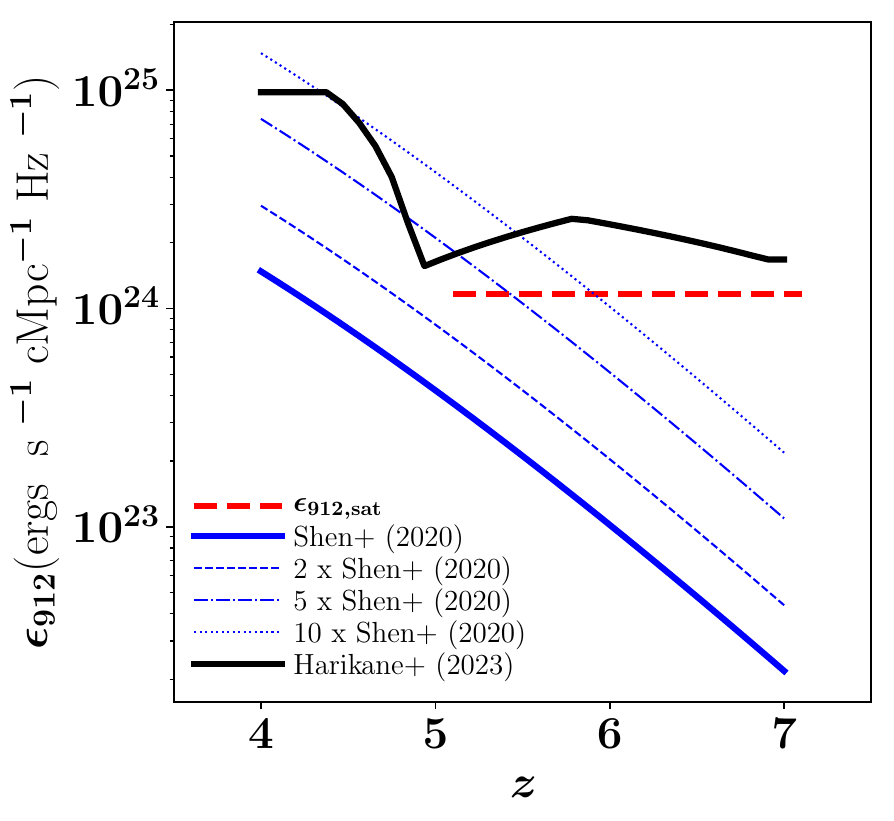}
\end{center}
\caption{The UV emissivity at 912 \AA\  (black jagged line) from the results of JWST \citep{harikane2023} compared to the fitting form of \citet[][blue solid line]{shen2020}. The dashed, dashed-dotted and dotted blue lines show the results of hiking the \citet{shen2020} emissivity by factors of $2 \times$, $5 \times$ and $10 \times$ respectively. The red dashed horizontal line denotes the constant value of comoving emissivity required \citep{haardt2015} to saturate the unresolved X-ray background measured by \citet{moretti2012}.}
\label{fig:epsilon912}
\end{figure}

We can use the measurements of \citet{harikane2023} { together with the inferences of \citet{haardt2015} to predict the  unresolved XRB from the JWST emissivity in the 912 \AA\ wave band. As an intermediate step, we compare the inferred emissivities $\epsilon_{912}$ from the results of the JWST \citep{harikane2023} to the best-fitting form of \citet{shen2020}:}
\begin{equation}
\epsilon_{912}(z) = \epsilon_0 (1+z)^a \frac{\exp(bz)}{\exp(cz) + d}
\end{equation}
where  $\epsilon_0 = 10^{24.108}$ ergs s$^{-1}$ Hz$^{-1}$ Mpc$^{-3}$, $a = 5.865$, $b = 0.731$, $c = 3.055$, $d = 15.60$.
This function is shown in the $z = \{4,7\}$ regime by the blue solid line  in Fig. \ref{fig:epsilon912}, compared to the JWST measurements shown by the black line.\footnote{Note that the \citet{shen2020} results assume a 100\% escape fraction, while those of \citet{harikane2023} assume 50\%. Hence, the results from \citet{harikane2023} have been scaled by factor 2 for comparison on the same plot.} { As can be seen from \eq{ixrb}, the integral under the emissivity curve, weighted by the redshift-dependent factor $c/[H(z)(1+z)^2]$ from \eq{reddep}, is proportional to the unresolved XRB implied by the corresponding UV measurement}. 

\citet{haardt2015} inverted the above analysis to set an upper limit on the (assumed constant for $z > 5$) comoving emissivity at 912 \AA\ which, along with an X-ray and UV spectral indices consistent with { current data \citep{ueda2014, masters2012} including corrections for obscured sources}, saturates the  unresolved XRB measured by \citet{moretti2012}.\footnote{The normalization of this unresolved  component [which is used to set the upper limit on the emissivity by \citet{haardt2015}] is found to be consistent with later measurements \citep[e.g.,][]{cappelluti2017}.} This is shown by the dashed red horizontal line in Fig. \ref{fig:epsilon912}.

{  Measuring  the area under the curve of the JWST emissivity (shown by the black jagged line in Fig. \ref{fig:epsilon912}), weighted by the redshift-dependent factor (see discussion following \eq{reddep}) and comparing it to that under the red dashed line weighted by the same factor, allows us to predict the enhancement in the unresolved XRB expected from the JWST}. This is shown by the black solid line in Fig. \ref{fig:xrb}, and found to be  about an order of magnitude higher than the constraints from using the {\it Swift} XRT  and {\it Chandra} observations  \citep{moretti2012}.

{ An estimate of the uncertainty in this result, shown by the gray shaded band, is made by invoking the scatter in the relation converting X-ray to UV luminosity \citep{shen2020}. The relation is usually fitted using two prefactors $\beta$ and $C$, 
with $\log {L_{\nu} (2 \ \rm keV) } = \beta \log {L_{\nu} (2500 \ {\rm \AA}) } + C$. The updated quasar observations are consistent with the prefactor values measured from \citet{steffen2006}, $\beta = 0.721 \pm 0.011$ and $C = 4.531 \pm 0.688$. The derived uncertainties are found to be consistent with the dispersion of the X-ray to UV ratio in quasars, constrained to $\sigma = 0.12$ \citep{chiaraluce2018}, of which about 56\% is contributed by  intrinsic variability.}

\begin{figure}
\begin{center}
\includegraphics[width =0.9\columnwidth]{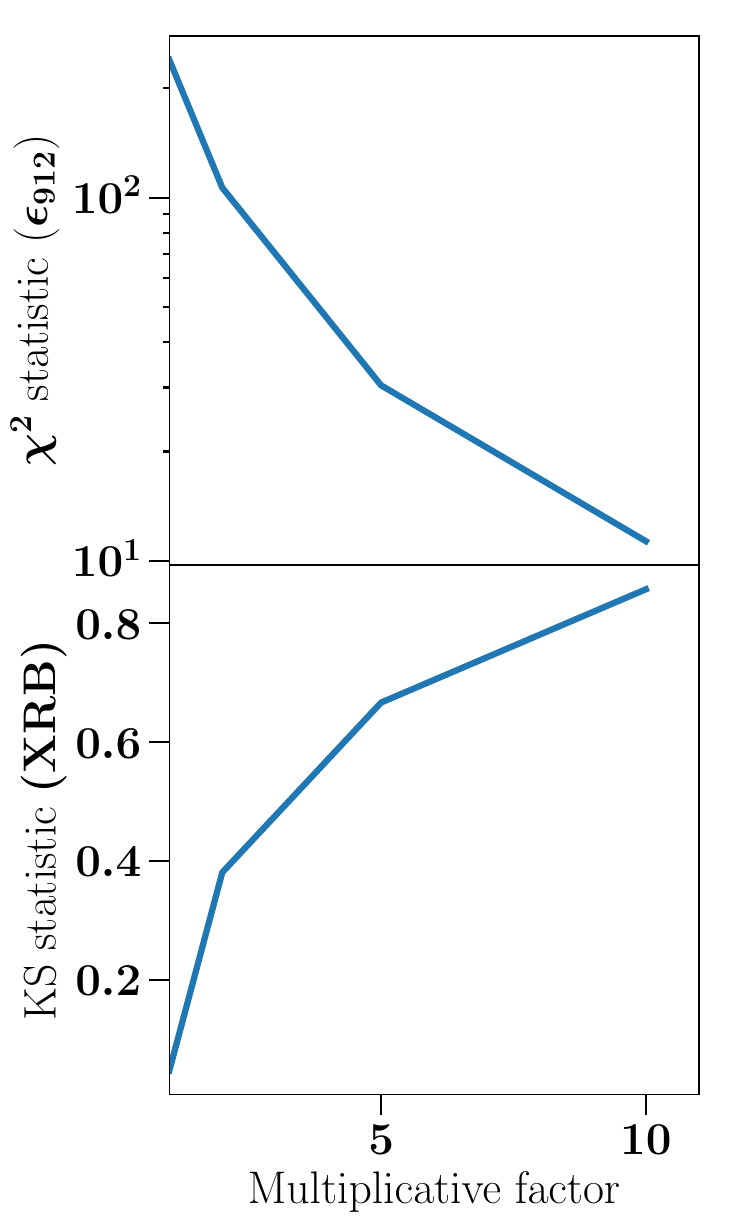} 
\end{center}
\caption{\textit{Top panel:} Value of the Cressie-Read $\chi^2$ statistic for the cases of multiplication of the emissivity at 912 \AA\ constrained by \citet{shen2020} by factors of $2 \times$, $5 \times$ and $10 \times$ respectively, compared to that measured by the JWST results \citep{harikane2023}. \textit{Lower panel:} The two-dimensional KS statistic ($d$-value), a measure of the deviation between the X-ray background data \citep{moretti2012} and that predicted by the \citet{shen2020} model multiplied by the same factors (higher value indicates larger deviation). As the multiplicative factor increases, the fit to the JWST $\epsilon_{912}$ becomes progressively better, with that to the XRB data becoming simultaneously worse.}
\label{fig:chisq}
\end{figure}

We can restate these findings in an equivalent manner by illustrating how the emissivity  at 912 \AA\ measured by JWST is discrepant with that required to saturate the current XRB limits. 
To quantify this discrepancy, { we use the 912 \AA\  emissivity inferred by \citet{shen2020}, shown by the solid blue line in Fig. \ref{fig:epsilon912}, as an intermediary to illustrate the scaling of the results.}  The values of the Cressie-Read $\chi^2$ statistic (and its associated $p$-value) are computed between the $\epsilon_{912, \rm JWST, i}$ and $\epsilon_{912, \rm S, i}$ quantities, which denote the `observed' \citep{harikane2023} and `expected' \citep{shen2020}  emissivities, and
where $i$ tracks redshift in the interval $z \in \{4,7\}$.
We also  measure the $\chi^2$ between the \citet{harikane2023} emissivities and those obtained by multiplying the \citet{shen2020} emissivities by factors of $2 \times$, $5 \times$ and $10 \times$ respectively (shown by the dashed, dashed-dotted and dotted blue lines respectively in Figure \ref{fig:epsilon912}). The value of the $\chi^2$ statistic for each case is plotted in the top panel of Figure \ref{fig:chisq}. Assuming a $p$-value of greater than $1\%$ for the curves to be deemed comparable, we find  that the $10 \times$ higher value comes closest to consistency with the JWST { emissivity curve} (with $p$-value 0.0101).

The comparison to the unresolved X-ray background data, however, runs in sharp contrast to the above behaviour for the emissivity. This is illustrated by predicting the XRB expected from the emissivity enhanced by the factors above { (by scaling the redshift-weighted areas under each of the blue curves to that of the red dashed line in Fig. \ref{fig:epsilon912}) and comparing the results to the prediction from the red dashed line, viz the \citet{moretti2012}} best-fitting curve in Fig. \ref{fig:xrb}. The expected backgrounds for the  $2 \times$, $5 \times$ and $10 \times$ multiplicative factors are illustrated by the dashed, dashed-dotted and dotted blue lines respectively in Figure \ref{fig:xrb}. The corresponding values of the two-dimensional Kolmogorov-Smirnov (KS) test statistic [$d$-value, which measures the deviation between the predicted XRB and that measured by \citet{moretti2012}] is plotted in the lower panel of Fig. \ref{fig:chisq} as a function of the multiplicative factor (a higher $d$-value indicates larger deviation, with a value of unity ruling out the model). The associated $p$-value of the 10$\times$ enhancement is $8.8 \times 10^{-8}$, reiterating the difficulty of simultaneously satisfying the JWST constraints on the 912 \AA\ emissivity over $z \sim 4-7$, and the XRB measurements.

\section{Discussion} 
We have explored the implications of the UV luminosity functions constrained by recent JWST AGN detections over $z \sim 4-7$, for the unresolved cosmic XRB. 
{ Using recent results connecting the UV emissivity to the intensity of the unresolved XRB}, we have found that the high number densities of quasars at the faint end of the luminosity function are likely disfavored by the XRB observations. The inferred XRB produced by this population would oversaturate the measured limits by about 1 order of magnitude. 

{ The approach of \citet{haardt2015} assumes an X-ray obscuration consistent with recent  findings \citep{ueda2014, masters2012}. For any level of dust obscuration, however, the UV (which is detected) is more heavily suppressed than the X-rays \citep[e.g.,][]{masters2012}. This implies that the observed UV luminosity function provides a lower limit on the X-ray luminosity function. The hard X-rays, originating in the compact corona of the AGN \citep[e.g.,][]{reis2013}, are also expected to be less collimated than the optical/UV that originates from the accretion disk. Thus, the results obtained here provide a conservative lower bound to the unresolved XRB expected from the JWST detections.\footnote{ A more distant explanation could be an intrinsically low X-ray/UV luminosity ratio in these objects (with an as-yet unknown physical cause), at a much higher level than the observationally-constrained scatter in the SED which is expected to be $\lesssim$ 0.2 dex \citep{chiaraluce2018,shen2020}.}}

The JWST UV LF is systematically higher than that measured at $z \sim 7$ with the SHELLQs survey \citep{matsuoka2023}.  At least 23 of the candidate AGN thus far have reported spectroscopic confirmations  with NIRSpec \citep{pacucci2023}. However, classical diagnostics alone have been deemed insufficient to confirm secure AGN detections since the NIRSpec resolutions are borderline ($R \sim 300$ compared to the required $R \sim 500$), which could also contribute to the source of the discrepancy \citep{ubler2023}. Other possibilities include biases from faint $H\alpha$ lines which, when corrected, may bring down the  number densities of the AGN by factors of $\sim 5$ \citep{matthee2023}.
Newer studies with JWST NIRSpec find comparable or higher UV LFs for the AGN at $z \sim 9$ than at $z \sim 6$ \citep{fujimoto2023}, with a 10-15\% AGN fraction in galaxies. 
Forthcoming spectroscopic census of AGN will help reveal the presence of other factors such as star-formation driven outflows that may make a dominant contribution to the luminosity \citep{matthee2023, zhang2023, ferrara2023} or objects such as brown dwarfs which may account for up to a third of the AGN candidate sample  \citep{greene2023, langeroodi2023}. 
  
\section*{Acknowledgements}  
HP's research is supported by the Swiss National Science Foundation via Ambizione Grant PZ00P2\_179934. The work of AL is supported in part by the Black Hole Initiative, which is funded by grants from the JTF and GBMF. We thank Richard Ellis, Andrea Ferrara and Pascal Oesch for comments on an earlier version of the manuscript, {and the referee for a helpful report.}

\def\aj{AJ}                   
\def\araa{ARA\&A}             
\def\apj{ApJ}                 
\def\apjl{ApJ}                
\def\apjs{ApJS}               
\def\ao{Appl.Optics}          
\def\apss{Ap\&SS}             
\def\aap{A\&A}                
\def\aapr{A\&A~Rev.}          
\def\aaps{A\&AS}              
\def\azh{AZh}                 
\def\baas{BAAS}
\def\jcap{JCAP}
\def\jrasc{JRASC}             
\def\memras{MmRAS}
\def\na{New Astronomy}
\def\nat{Nature}
\def\mnras{MNRAS}             
\def\pra{Phys.Rev.A}          
\def\prb{Phys.Rev.B}          
\def\prc{Phys.Rev.C}          
\def\prd{Phys.Rev.D}          
\def\prl{Phys.Rev.Lett}       
\def\pasp{PASP}               
\def\pasj{PASJ}
\def\physrep{Phys. Repts.}
\def\qjras{QJRAS}             
\def\skytel{S\&T}             
\def\solphys{Solar~Phys.}     
\def\sovast{Soviet~Ast.}      
\def\ssr{Space~Sci.Rev.}      
\def\zap{ZAp}                 
\let\astap=\aap
\let\apjlett=\apjl
\let\apjsupp=\apjs

\small{
\bibliographystyle{plainnat}
\bibliography{mybib}
}

\end{document}